\documentclass[12pt]{article}
\newcommand{\nc}{\newcommand}
\usepackage{amsfonts}
\usepackage{amssymb}
\usepackage{amsbsy}
\nc{\HH}{\mathbb{H}}
\nc{\bib}{\bibitem}
\nc{\al}{\alpha}
\nc{\g}{\gamma}
\nc{\G}{\Gamma}
\nc{\D}{\Delta}
\nc{\eps}{\epsilon}
\nc{\la}{\lambda}
\nc{\La}{\Lambda}
\nc{\var}{\varphi}
\nc{\pa}{\partial}
\nc{\nn}{\nonumber \\ }
\nc{\hf}{\frac{1}{2}}  
\nc{\dz}{\frac{dz}{2\pi i}}
\nc{\bin}[2]{\left (\begin{array}{c} {#1}\\ {#2} \end{array}\right )}
\nc{\ben}{\begin{equation}}
\nc{\een}{\end{equation}}
\nc{\bea}{\begin{eqnarray}}
\nc{\eea}{\end{eqnarray}}
\nc{\bra}[1]{\langle {#1}|}
\nc{\ket}[1]{|{#1}\rangle}
\nc{\C}{\mbox{\hspace{1.24mm}\rule{0.2mm}{2.5mm}\hspace{-2.7mm} C}}
\nc{\Nat}{\mbox{\hspace{.04mm}\rule{0.2mm}{2.8mm}\hspace{-1.5mm} N}}
\newcommand{\R}{\mbox{\hspace{.04mm}\rule{0.2mm}{2.8mm}\hspace{-1.5mm} R}}
\def\vvdots{\mathinner{\mkern1mu\raise1pt\vbox{\kern7pt\hbox{.}}\mkern2mu
 \raise4pt\hbox{.}\mkern2mu\raise7pt\hbox{.}\mkern1mu}}

\def\min{{\rm min}}

\begin{document}

\topmargin -5mm
\oddsidemargin 5mm

\begin{titlepage}
\setcounter{page}{0}

\vspace{8mm}
\begin{center}
{\huge SLE-type growth processes and the}\\[.4cm]
{\huge Yang-Lee singularity}

\vspace{15mm}
{\Large Fr\'ed\'eric Lesage}\ \  and\ \  {\Large J{\o}rgen Rasmussen}\\[.3cm] 
{\em Centre de recherches math\'ematiques, Universit\'e de Montr\'eal}\\ 
{\em Case postale 6128, 
succursale centre-ville, Montr\'eal, Qc, Canada H3C 3J7}\\[.3cm]
lesage,rasmusse@crm.umontreal.ca

\end{center}

\vspace{10mm}
\centerline{{\bf{Abstract}}}
\vskip.4cm
\noindent
The recently introduced SLE growth processes are based on conformal maps
from an open and simply-connected 
subset of the upper half-plane to the half-plane itself.
We generalize this by considering a hierarchy of stochastic evolutions
mapping open and simply-connected
subsets of smaller and smaller fractions of the upper
half-plane to these fractions themselves. The evolutions are all driven by
one-dimensional Brownian motion. Ordinary chordal SLE appears at grade one
in the hierarchy. At grade two we find a direct correspondence
to conformal field theory through the explicit construction
of a level-four null vector in a highest-weight module of the Virasoro 
algebra. This conformal field theory has central charge $c=-22/5$ and is
associated to the Yang-Lee singularity. Our construction may thus offer a
novel description of this statistical model.
\\[.5cm]
{\bf Keywords:} Stochastic L\"owner evolution, 
 conformal field theory, Yang-Lee singularity. 
\end{titlepage}
\newpage
\renewcommand{\thefootnote}{\arabic{footnote}}
\setcounter{footnote}{0}

\section{Introduction}

A new approach to the description of conformal field theories (CFTs) in two
dimensions has recently appeared where instead of discussing objects
in terms of local fields and their fusions, one is rather interested
in a description based on spatially extended quantities defined through
geometry. The differential equations of
the stochastic L\"owner evolution (SLE) have emerged as a
mathematically precise way of describing certain CFTs directly in the
continuum, without reference to an underlying lattice.

The chordal SLE processes are constructed through conformal maps
from a subset of the upper half-plane onto the half-plane itself.
The processes are driven by the random one-dimensional Brownian
motion. Properties thereby described have an intrinsic geometrical
nature.

The study of these stochastic evolutions or growth processes was initiated
by Schramm \cite{Sch} and has been pursued further in 
\cite{Wer,LSW01,LSW01a,Smi,LSW01b,LSW01c,RS,LSW0209,FW}, for example. 
A review for physicists may be found in \cite{Car}, while
\cite{Law} contains a mathematical introduction.

An explicit relationship between SLE and CFT has been elucidated
recently \cite{BB1,BB2} by considering random walks on the Virasoro
group. The link is found through a singular vector at level two
in highest-weight modules. The kernel of the vector
corresponds to conserved quantities under the random process.

Although the correspondence exists, the number of CFTs having
geometrical properties described by SLE is still very limited.
Furthermore, there is no apparent pattern assisting in the
identification of these new descriptions of field theories.

The aim of the present work is to show that there might be 
conformal systems described by generalizations of SLE.
The approach of Bauer and Bernard \cite{BB1,BB2} 
may be extended to more general walks than the
one generating SLE. A particular class of extensions
corresponds to a hierarchy of stochastic evolutions in which
SLE appears at grade one. 
These growth processes are associated to conformal maps of 
open and simply-connected
subsets of smaller and smaller fractions of the upper half-plane onto
the fractions themselves, and are all driven by one-dimensional
Brownian motion.
Using two-sided Brownian motion the stochastic processes may
be extended to also describing flows from fractions of the
upper half-plane to subsets thereof.
At grade two in the hierarchy we find a link to the Yang-Lee singularity
through the construction of a level-four null vector.
This in turn potentially offers a new geometrical description of that 
statistical model.

\section{Stochastic evolutions}

\subsection{Stochastic L\"owner evolution}

Let $Y_t$ be a real-valued continuous function, $t\geq0$.
For each element in the upper half-plane, $z\in\HH$,
we consider the solution $g_t(z)$ to L\"owner's differential equation
\ben
 \pa_tg_t(z)\ =\ \frac{2}{g_t(z)-Y_t}\ ,\ \ \ \ 
  \ \ \ \ \ g_0(z)\ =\ z
\label{L}
\een
The factor 2 is conventional but could be changed by re-normalization.
Let $\tau=\tau(z)$ denote the time such that the solution
$g_t(z)$ exists for all $t\in[0,\tau]$, while for increasing time 
$\lim_{t\to\tau}g_t(z)=Y_\tau$. Following \cite{Sch,RS,Law}, 
one may define the evolving hull $K_t$ as the closure of
$\{z\in\HH:\tau(z)\leq t\}$.
In time, it is an increasing sequence of compact sets.
As a conformal map from the simply-connected domain
$\HH\setminus K_t$ onto the open half-plane $\HH$,
$g_t$ is uniquely determined by the so-called
hydrodynamic normalization at infinity:
\ben
 \lim_{z\to\infty}\left(g_t(z)-z\right)\ =\ 0
\label{hydro}
\een

Stochastic L\"owner evolutions are growth processes defined by
choosing standard one-dimensional (and one-sided)
Brownian motion, $B_t$, as
the driving function: $Y_t=\sqrt{\kappa}B_t$, with $B_0=0$. The parameter
$\kappa$ characterizes the process which is denoted 
SLE$_\kappa$. For $t,s\geq0$, the expectation value is normalized as
${\bf E}[(\sqrt{\kappa}B_t)(\sqrt{\kappa}B_s)]=\kappa\ \min(t,s)$.

One defines the function 
\ben
 f_t(z):=\ g_t(z)-Y_t
\een
It follows that it satisfies the differential equation
\ben
 \pa_tf_t(z)\ =\ \frac{2}{f_t(z)}-\pa_tY_t\ ,\ \ \ \ 
  \ \ \ \ \ f_0(z)\ =\ z
\label{f}
\een
When $Y_t$ denotes Brownian motion its
time derivative is thought of as white noise: $dB_t/dt\sim W_t$. 
The inverse of the function $f_t$ is related to the inverse of the SLE map:
$f^{-1}_t(z)=g^{-1}_t(z+Y_t)$. The trace $\gamma$ of SLE
is then defined by
\ben
 \gamma(t):=\ \lim_{z\to0}f_t^{-1}(z)
\label{gamma}
\een 
By construction, $z$ is an element of $\HH$, so the limit
is taken from the upper half-plane only. 
The nature of the trace is known to depend radically on $\kappa$
\cite{RS}:
for $0\leq\kappa\leq4$ it is a simple curve, for $4<\kappa<8$
a self-intersecting curve, whereas for $8<\kappa$ it is space filling.
The Hausdorff dimension of the SLE$_\kappa$ trace is discussed
in \cite{RS,Bef1,Bef2}.

\subsection{Hierarchy of stochastic evolutions}

For positive integer $n$, we define the open subset of $\HH$
\ben
 \HH_n\ =\ \{z\in\HH: z=re^{i\theta};\ r\in\R_>;\ 0<\theta<\pi/n\}
\label{Hn}
\een
Note that $\HH_1$ is the upper half-plane itself.
We now introduce a hierarchy of L\"owner-like differential equations 
whose solutions have properties similar to the SLE maps.
For positive integer $n$ we define the differential equation
\ben
 \pa_tg_t(z)\ =\ \frac{2}{g_t^{n-1}(z)\left(g_t^n(z)-Y_t\right)}\ ,
  \ \ \ \ \ \ \ \ \ g_0(z)\ =\ z
\label{hier}
\een
with $Y_0=0$. For each $z\in\HH_n$ the solution is well-defined up to
a time $\tau_n(z)$. Similarly to the ordinary SLE case, 
the differential equation (\ref{hier}) describes the evolution of the hull
$K^{(n)}_t$ defined  as the closure of $\{z\in\HH_n: \tau_n(z)\leq t\}$.

The solution to (\ref{hier}) is a conformal map from 
$\HH_n\setminus K^{(n)}_t$ onto $\HH_n$. To see this, one may
generalize the proof of Proposition 2.2 in \cite{Law}.
One first verifies that $\pa_zg_t(z)$ is well-defined by analyzing
$\pa_t\pa_zg_t(z)$. From the evaluation of $\pa_t(g_t(z)-g_t(z'))$
(which is shown to have $(g_t(z)-g_t(z'))$ as a factor), one deduces
that $g_t(z)\neq g_t(z')$ when $z\neq z'$. It has thereby
been established that $g_t$ is a conformal transformation of 
$\HH_n\setminus K_t^{(n)}$. To show that $g_t(\HH_n\setminus K_t^{(n)})
=\HH_n$, one studies the inverse flow $h_t(w)$, $w\in\HH_n$, which
is a solution to
\ben
 \pa_th_t(w)\ =\ -\frac{2}{h_t^{n-1}(w)(h_t^n(w)-Y_{t_0-t})}\ ,\ \ \ 
  \ \ \ \ h_0(w)\ =\ w
\een
for some $t_0\geq0$. The solution $h_t(w)$ is well-defined for 
$0\leq t\leq t_0$ since $\pa_t{\rm Im}(h_t^n(w))>0$ and $|h_t^{n-1}|\geq
\min\{|h_t^n(w)|,1\}$. This ensures that the solution cannot hit the
singularities. With $z=h_{t_0}(w)$, $g_t(z)=h_{t_0-t}(w)$ 
is seen to be a solution to (\ref{hier}) (implying that $h_t(w)$
is indeed the inverse flow), and $g_{t_0}(z)=w$ showing that
$w\in g_{t_0}(\HH_n\setminus K_t^{(n)})$.

The solution to (\ref{hier}) is determined uniquely by
the hydrodynamic normalization at infinity (\ref{hydro}). It has
the power series expansion
\ben
 g_t(z)\ =\ z+\frac{2t}{z^{2n-1}}+{\cal O}(1/|z|^{(2n)})\ ,
  \ \ \ \ \ \ z\to\infty
\label{power}
\een
We refer to the process as being of grade $n$.

When $n=1$ (and $Y_t=\sqrt{\kappa}B_t$) we recover the ordinary SLE 
equation (\ref{L}). In a subsequent section we shall focus on grade 
$n=2$ as it is in this case we find a new relation to CFT and
the Yang-Lee singularity.

Two important properties of ordinary SLE are scale invariance
and a sort of stationarity. These apply to solutions to
(\ref{hier}) as well. In the spirit of Proposition 2.1 in \cite{RS}
(see also \cite{LSW01}), we have that 
the growth process defined by (\ref{hier}) is scale invariant in the 
following sense. For $\al>0$ the process
$t\mapsto\al^{-1/(2n)}K^{(n)}_{\al t}$ has the same law as 
$t\mapsto K^{(n)}_{t}$, while the process $(z,t)\mapsto\al^{-1/(2n)}
g_{\al t}(\al^{1/(2n)}z)$ has the same law as $(z,t)\mapsto g_t(z)$.
Also, the map $\tilde{g}(z):=(g_{t_1}\circ g^{-1}_{t_0})(z+Y_{t_0})-Y_{t_0}$
has the same law as $g_{t_1-t_0}$ when $t_1>t_0>0$.
Moreover, $\tilde{g}$ is independent of $g_{t_0}$.
These assertions can be proved by a simple adaptation of the proof for
ordinary SLE.

We define $f_t$ through
\ben
 f^n_t(z)\ =\ g^n_t(z)-Y_t
\label{fn}
\een
It follows that $f_t(z)$ satisfies the differential equation
\ben
 \pa_tf_t(z)\ =\ \frac{2}{f^{2n-1}_t(z)}-\frac{1/n}{f^{n-1}_t(z)}\pa_tY_t\
  ,\ \ \ \ \ \ \ \ f_0(z)\ =\ z
\label{dfn}
\een
with a canonical choice of boundary condition. The solution
respects the hydrodynamic
normalization at infinity (\ref{hydro}), and 
it corresponds to choosing the 'principal root' in the
relation (\ref{fn}). As in ordinary SLE, 
we use $f_t$ to define an SLE-type trace for the hierarchy of evolutions:
\ben
 \gamma_n(t):=\ \lim_{z\to0}f^{-1}_t(z)
\een 

To illustrate our construction, we now consider the situation where
the driving function vanishes for all $t$: $Y_t\equiv0$ (corresponding to
$\kappa\equiv0$). The differential equation becomes
\ben
 \pa_tg_t(z)\ =\ \frac{2}{g_t^{2n-1}(z)}\ ,\ \ \ \ \ \ \ \ \ \ g_0(z)\ =\ z
\een
with solution
\ben
 g_t(z)\ =\ \left(z^{2n}+4nt\right)^{1/(2n)}
\een
The trace reads
\ben
 \gamma_n(t)\ =\ |(4nt)^{1/(2n)}|e^{i\pi/(2n)}
\een
while the hull is
\ben
 K^{(n)}_t\ =\ \{re^{i\pi/(2n)}: r\in[0,|(4nt)^{1/(2n)}|]\}
\een

Following \cite{RS} on ordinary SLE, we may take $B$ to be two-sided 
Brownian motion (or more generally, $Y$ to be defined for negative $t$
as well). The equation (\ref{hier}) can then also be solved for
negative $t$, in which case $g_t$ is a conformal map from $\HH_n$ into
a subset of $\HH_n$. Indeed, Lemma 3.1 in \cite{RS} extends to our case.
In the extended version it states that the map $z\mapsto g_{-t}(z)$ has the
same distribution as the mapping of $z$ into the principal $n$th root
of  $\left(\left(g_t^{-1}((z^n+Y_t)^{1/n})\right)^n-Y_t\right)$. To see this
we first observe that for $z\in\HH_n$ the principal $n$th root of
$(z^n+x)$ for $x$ real also lies in $\HH_n$. For $t_1\in\R$
we then define the
function $\hat{g}_t^{(t_1)}$ as the principal root in the functional relation
\ben
 \left(\hat{g}_t^{(t_1)}(z)\right)^n\ =\ 
   \left(g_{t_1+t}\circ g_{t_1}^{-1}((z^n+Y_{t_1})^{1/n})\right)^n-Y_{t_1}
\label{ghat}
\een
It follows that $\hat{g}_t^{(t_1)}(z)$ is a solution to 
\ben
 \pa_t\hat{g}_t^{(t_1)}(z)\ =\ 
  \frac{2}{\left((\hat{g}_t^{(t_1)}(z)\right)^{n-1}
   \left(\left((\hat{g}_t^{(t_1)}(z)\right)^n
    -(Y_{t_1+t}-Y_{t_1})\right)} \ ,\ \ \ \ \ \ \hat{g}_0^{(t_1)}(z)\ =\ z
\label{diffghat}
\een
We note that $\hat{Y}_t^{(t_1)}:=Y_{t_1+t}-Y_{t_1}$ has the same law as
$Y_t$ as maps from $\R$ to $\R$, and since 
\ben
 \left(\hat{g}_{-t_1}^{(t_1)}(z)\right)^n\ =\ 
   \left(g_{t_1}^{-1}((z^n+Y_{t_1})^{1/n})\right)^n-Y_{t_1}
\een
the assertion of the extended lemma follows.

With two-sided Brownian motion at hand, we may define alternatively
to (\ref{fn})
\ben
 f^n_t(z)\ =\ g^n_{-t}(z)-Y_{-t}
\label{fna}
\een
satisfying
\ben
 \pa_tf_t(z)\ =\ \frac{-2}{f^{2n-1}_t(z)}
  -\frac{1/n}{f^{n-1}_t(z)}\pa_tY_{-t}\
  ,\ \ \ \ \ \ \ \ f_0(z)\ =\ z
\label{dfna}
\een
Choosing the driving function as $Y_t=\sqrt{\kappa}B_{-t}$ we have
\ben
 df_t(z)\ =\ \frac{-2}{f^{2n-1}_t(z)}dt
  -\frac{\sqrt{\kappa}/n}{f^{n-1}_t(z)}dB_{t}\
  ,\ \ \ \ \ \ \ \ f_0(z)\ =\ z
\label{dfnaB}
\een
This will appear in the link to the Yang-Lee singularity addressed below.

It is remarked that our construction may be interpreted as chordal SLE
in $\HH_n$. To appreciate this, we introduce $\phi_n(z)=z^n$ and let
$G_t^{(n)}(z)$ denote the map in (\ref{hier}) when $Y_t=\sqrt{\kappa}B_t$.
$G_t^{(1)}(z)$ thus
corresponds to chordal SLE in the upper half-plane $\HH_1=\HH$.
We then have that
\ben
 \hat{G}_t^{(n)}(z)\ =\ \phi_n^{-1}\circ G_{nt}^{(1)}\circ\phi_n(z)
\een
(where $\phi_n^{-1}$ singles out the principal root) satisfies
(\ref{hier}) albeit with $\hat{\kappa}=n\kappa$.
As a consequence, $K_t^{(n)}$ is seen to correspond
to $\phi_n^{-1}(K_{nt}^{(1)})$.

\section{Relation to conformal field theory}

\subsection{Ordinary SLE}

Bauer and Bernard \cite{BB1,BB2} have recently discussed a direct 
relationship between SLE$_\kappa$ and CFT.
Their construction starts from a random walk on the (somewhat formal)
Virasoro group:
\ben
 G_t^{-1}dG_t\ =\ -2L_{-2}dt+\sqrt{\kappa}L_{-1}\circ
  dB_t\ ,\ \ \ \ \ \ \ G_0=1
\een
here written in the Stratonovich interpretation. We shall
rather discuss it in the Ito form where it reads
\ben
 G_t^{-1}dG_t\ =\ \left(-2L_{-2}+\frac{\kappa}{2}L_{-1}^2\right)dt
  +\sqrt{\kappa}L_{-1}dB_t\ ,\ \ \ \ \ \ \ G_0=1
\label{ItoG}
\een
$G_t$ is an element of $Vir_-$ obtained by exponentiating the negative
modes, $L_n$, $n<0$, of the Virasoro algebra. We write a generic element
$G\in Vir_-$ as
\ben
 G\ =\ \cdots e^{x_2L_{-2}}e^{x_1L_{-1}}
\label{GGBB}
\een
and recall the definition of the Virasoro algebra:
\ben
 [L_n,L_m]\ =\ (n-m)L_{n+m}+ \frac{c}{12}n (n^2-1) \delta_{n+m,0}
\label{vir}
\een
The central charge $c$ plays a prominent role in CFT.
As we shall discuss, it is through the
construction of singular vectors in highest-weight modules of the
Virasoro algebra that the connection to SLE$_\kappa$ is established
\cite{BB1,BB2}.

The conformal transformation generated by (\ref{ItoG}) acts on a primary
field of weight $\Delta$ as
\ben
 G_t^{-1}\phi_\Delta (z) G_t\ 
  =\ \left(\partial_z f_t(z)\right)^\Delta \phi_\Delta (f_t(z))
\label{GphiG}
\een
for some conformal map $f_t$ to be determined\footnote{For simplicity,
we do not distinguish explicitly between boundary and bulk primary
fields, nor do we write the anti-holomorphic part.}.
Using that the Virasoro generators act as
\ben
 [L_n,\phi_\Delta(z)]\ =\ (z^{n+1}\partial_z + \Delta(n+1)z^n) \phi_\Delta(z)
\label{Lphi}
\een
one finds that the conformal map associated to the random process 
(\ref{ItoG}) must be a solution to the stochastic differential equation
\ben
 df_t(z)\ =\ \frac{2}{f_t(z)}dt-\sqrt{\kappa}dB_t\ ,\ \ \ \ 
  \ \ f_0(z)\ =\ z
\label{dfdt}
\een
corresponding to
(\ref{f}). This follows from computing the Ito differential of (\ref{GphiG})
and is discussed in more details in \cite{BB1,BB2}.

Observables of the random process (\ref{ItoG}) are thought of as functions 
on the Virasoro group, $F(G_t)$, and a goal is to find
the evolution for the expectation values of these observables.
With the left-invariant  vector fields, $\nabla_\ell$, defined by 
\ben
 \nabla_{\ell}F(G_t)\ =\ \frac{d}{du} F(G_te^{uL_\ell})|_{u=0}
\label{u}
\een
one has \cite{BB1} that the expectation value ${\bf E}[F(G_t)]$ satisfies 
\ben
 \partial_t {\bf E}[F(G_t)]\ =\ 
  {\bf E}[\left(-2\nabla_{-2}+\frac{\kappa}{2}\nabla_{-1}^2\right)F(G_t)]
\label{dE}
\een

We shall be interested in observables of the form
$F_\Delta(G_t)=G_t\ket{\Delta}$ where
$\ket{\Delta}$ is the highest-weight vector of weight $\Delta$
in the Verma module ${\cal V}_\Delta=Vir_-\ket{\Delta}$
(see \cite{BB1,BB2} and below).
In this case the expectation value reads
\ben
 \pa_t{\bf E}[G_t\ket{\Delta}]\ =\ {\bf E}[G_t\left(-2L_{-2}
  +\frac{\kappa}{2}L_{-1}^2\right)\ket{\Delta}]
\label{LD}
\een
For some values of $\kappa$ (in relation to the central charge $c$),
the linear combination $-2L_{-2}+\frac{\kappa}{2}L_{-1}^2$ will
produce a singular vector when acting on the highest-weight vector in a
highest-weight module. This is an important point as it enables one, 
through the representation theory of the algebra, to find quantities 
conserved in mean under the random process.

The Verma module ${\cal V}_\Delta$ contains the singular vector at level two
\ben
 \ket{\Delta;2}\ =\ \left(L_{-2}-\frac{\kappa}{4}L_{-1}^2\right)\ket{\Delta}
\label{d2}
\een
provided $L_1\ket{\Delta;2}=L_2\ket{\Delta;2}=0$. It is straightforward
to show that this implies the parameterizations
\ben
 c_\kappa\ =\ 1-\frac{3(4-\kappa)^2}{2\kappa}\ ,\ \ \ \ \ \ 
 \Delta_\kappa\ =\ \frac{6-\kappa}{2\kappa}
\label{cd}
\een
The expectation value of the observable $F_\Delta(G_t)=G_t\ket{\Delta}$
thus vanishes (\ref{LD}):
\ben
 \pa_t{\bf E}[G_t\ket{\Delta}]\ =\ 0
\label{EG0}
\een
We see that this direct relationship between SLE$_\kappa$ evolutions
and CFT is through the existence of a level-two
singular vector in a highest-weight module. As discussed in \cite{BB2},
this relationship provides links between conformal correlation
functions and probabilities in SLE$_\kappa$.

\subsection{Extended SLE}

Since Brownian motion played a significant role in the derivation
of (\ref{dE}) and (\ref{LD}), and hence in the correspondence between 
SLE and CFT, it remains unclear how to treat more general random 
processes than (\ref{ItoG}).
An extension invites itself, though. Namely, consider the random walk
\ben
 G_t^{-1}dG_t\ =\ v_{-n}L_{-2n}dt
  +\sqrt{\kappa}u_{-n}L_{-n}\circ dB_t\ ,\ \ \ \ \ \ \ G_0\ =\ 1
\label{2nS}
\een
or in the Ito interpretation
\ben
 G_t^{-1}dG_t\ =\ \left(v_{-n}L_{-2n}+\frac{\kappa u_{-n}^2}{2}
  L_{-n}^2\right)dt
  +\sqrt{\kappa}u_{-n}L_{-n}dB_t\ ,\ \ \ \ \ \ \ G_0\ =\ 1 
\label{2nI}
\een
In this case we have
\ben
 \partial_t {\bf E}[F(G_t)]\ =\ 
  {\bf E}[\left(v_{-n}\nabla_{-2n}+
  \frac{\kappa u_{-n}^2}{2}\nabla_{-n}^2\right)F(G_t)]
\label{dEn}
\een
and in particular
\ben
 \partial_t {\bf E}[G_t\ket{\D}]\ =\ 
  {\bf E}[G_t\left(v_{-n}L_{-2n}+\frac{\kappa u_{-n}^2}{2}
  L_{-n}^2\right)\ket{\D}]
\label{EF2n}
\een
To relate this to the SLE-type differential equations discussed above,
we write (\ref{dfn}) and (\ref{dfnaB}) uniformly as
\ben
 df_t(z)\ =\ \frac{2(-1)^{s+1}}{(f_t(z))^{2n-1}}dt-\frac{\sqrt{\kappa}/n}{
  (f_t(z))^{n-1}}dB_t
\label{s}
\een
where $s=1$ or $s=2$ depending on the choice of relation
(\ref{fn}) or (\ref{fna}), respectively.
Taking the Ito differential of the right hand side of (\ref{GphiG})
results in
\bea
 &&d\{(\pa_z f_t(z))^\D\phi_\D(f_t(z))\}\nn
&=&
  \left[\left(2(-1)^{s+1}+\frac{\kappa(n-1)}{2n^2}\right)L_{-2n}dt
   -\frac{\sqrt{\kappa}}{n}L_{-n}dB_t,(\pa_z f_t(z))^\D\phi_\D(f_t(z))\right]
  \nn
 &&+\frac{\kappa}{2n^2}\left[L_{-n},\left[L_{-n},
  (\pa_z f_t(z))^\D\phi_\D(f_t(z))\right]\right]dt
\label{dRHS}
\eea
while the Ito differential of the left hand side of (\ref{GphiG}) 
generated by the random walk (\ref{2nI}) reads
\bea
 d\{G_t^{-1}\phi_\D(f_t(z))G_t\}&=&\left[-v_{-n}L_{-2n}dt
  -\sqrt{\kappa}u_{-n}L_{-n}dB_t,G_t^{-1}\phi_\D(f_t(z))G_t\right]\nn
 &&+\frac{\kappa u_{-n}^2}{2}\left[L_{-n},\left[L_{-n},
   G_t^{-1}\phi_\D(f_t(z))G_t\right]\right]dt
\label{dGphiG}
\eea
A comparison of the two Ito differentials 
suggests considering the walk
\ben
 G_t^{-1}dG_t\ =\ \left(\left(2(-1)^s-\frac{\kappa(n-1)}{2n^2}\right)L_{-2n}
  +\frac{\kappa}{2n^2}L_{-n}^2\right)dt+\frac{\sqrt{\kappa}}{n}L_{-n}dB_t
\label{GdGdtdB}
\een

According to this, 
we should be looking for singular vectors of the form
\ben
 \ket{\Delta;2n}\ =\ \left(\left(2(-1)^s-
  \frac{\kappa(n-1)}{2n^2}\right)L_{-2n}
  +\frac{\kappa}{2n^2}L_{-n}^2\right)\ket{\Delta}
\label{d2n}
\een
The upset, however, is that for $n>1$
\bea
 L_1\ket{\D;2n}
 &=&\left(\left((2n+1)\left(2(-1)^s
  -\frac{\kappa(n-1)}{2n^2}\right)+\frac{(n+1)\kappa}{2n^2}\right)L_{-(2n-1)}
  \right.\nn
 &&\ \ \ \ \ \ \ \ \left.
   +\frac{(n+1)\kappa}{n^2}L_{-n}L_{-(n-1)}\right)\ket{\D}\nn
 &\neq&0
\label{L1}
\eea
for all $\kappa$. This means that (\ref{d2n}) can be a singular vector only
when $n=1$, and is then given by (\ref{d2}) (when $s=1$).
One should not be completely discouraged by this. The pivotal property of 
the state $\left(L_{-2}-\frac{\kappa}{4}L_{-1}^2\right)\ket{\Delta}$ 
appearing in (\ref{LD}) and
applications thereof \cite{BB2}, is that it vanishes in the
quotient space of ${\cal V}_\Delta$ where all singular vectors have
been factored out. In other words, it is a null vector.
This means that we do not have to insist that
the vector (\ref{d2n}) is a (primitive) singular vector itself, but only 
require that it is a linear combination of descendants of (primitive)
singular vectors. An example is provided below.

\section{Yang-Lee singularity}

Generically, the Verma module 
${\cal V}_\Delta$ is irreducible. Minimal models \cite{BPZ,bible}
are examples of CFTs for which it is reducible.
They are labeled by a pair of positive and co-prime integers
$p>p'$, and are denoted ${\cal M}(p,p')$. The central charge is
\ben
 c\ =\ 1-6\frac{(p-p')^2}{pp'}
\label{cmm}
\een
while the spectrum of primary fields or highest-weight representations
have conformal weights
\ben
 \D_{r,s}\ =\ \frac{(rp-sp')^2-(p-p')^2}{4pp'}\ ,
  \ \ \ \ 1\leq r<p'\ ,\ \ \ \ 1\leq s<p
\label{Dmm}
\een
with $\D_{p'-r,p-s}=\D_{r,s}$.
There are two singular vectors not being
descendants of singular vectors themselves, and they appear at
levels $rs$ and $(p'-r)(p-s)$. 

For $p'=2$, there is only one primary field admitting a singular vector
at level two, in which case $(r,s)=(1,2)$. For $p'>2$, on the other hand, 
there are two such fields, labeled by $(1,2)$ and $(2,1)$, respectively.
It is easily verified that 
\ben
 \D_{1,2}\ =\ \D_{\kappa=4p/p'},\ \ \ \ \ \ \D_{2,1}\ =\ \D_{\kappa=4p'/p}
 \een
It follows that SLE$_\kappa$ and SLE$_{16/\kappa}$, with
$\kappa=4p/p'$, may be linked to
the same minimal model, albeit via two different primary fields in the model.

The simplest example of a null vector of the form (\ref{d2n}) for $n>1$
that we have found is a level-four vector in the minimal 
model ${\cal M}(5,2)$ with $c=-22/5$, cf. (\ref{cmm}). This model offers
a CFT description of the statistical Yang-Lee singularity.
Unlike ordinary SLE (except SLE$_{\kappa=6}$ which is known to correspond to 
percolation, and has central charge $c=0$), the null vector appears in the 
{\em identity module} having singular vectors
\bea
 \ket{0}_1&=&L_{-1}\ket{0}\nn 
 \ket{0}_4&=&\left(L_{-4}+\frac{5}{27}L_{-3}L_{-1}-\frac{5}{3}L_{-2}^2
  +\frac{125}{27}L_{-2}L_{-1}^2-\frac{125}{108}L_{-1}^4\right)\ket{0}
\label{14}
\eea
The null vector of our interest reads
\bea
 \ket{0;4}&=&\ket{0}_4+\left(-\frac{5}{27}L_{-3}-\frac{125}{27}L_{-2}L_{-1}
  +\frac{125}{108}L_{-1}^3\right)\ket{0}_1\nn
 &=&\left(L_{-4}-\frac{5}{3}L_{-2}^2\right)\ket{0}
\label{nullYL}
\eea
Comparing this to (\ref{d2n}), we find that the Yang-Lee singularity is
related to a grade-two SLE-type evolution with 
\ben
 \kappa\ =\ 40
\label{kappa}
\een
and $s=2$. There is no non-negative solution for $\kappa$ 
when $s=1$.
In summary, this SLE-type evolution reads 
\ben
 \pa_tg_t(z)\ =\ \frac{2}{g_t(z)\left(g_t^2(z)-\sqrt{40}B_t\right)}\ ,
  \ \ \ \ \ \ \ g_0(z)\ =\ z
\label{dg2}
\een
or in terms of $f_t(z)$, cf. (\ref{dfnaB}) and (\ref{s}):
\ben
 df_t(z)\ =\ \frac{-2}{f_t^3(z)}dt
  -\frac{\sqrt{10}}{f_t(z)}dB_t\ ,
  \ \ \ \ \ \ \ f_0(z)\ =\ z
\label{df2}
\een
This provides a novel approach to the Yang-Lee model, and may
eventually lead to an explicit geometric realization.

\section{Conclusion}

As observed by Bauer and Bernard \cite{BB1,BB2}, SLE may be linked
to CFT through the construction of a singular vector
in a Virasoro highest-weight module. We have extended their approach,
and found that the Yang-Lee singularity may be described by
a generalization of the SLE differential equation.
This new stochastic evolution appears at grade two in a hierarchy of 
SLE-type growth processes in which ordinary SLE appears at grade one.
Their approach has recently been extended also to stochastic evolutions
in superspace and superconformal field theory \cite{Ras}.
Another extension will appear elsewhere  
where it is discussed how SLE-type growth processes may be linked
to CFT via (non-primary) descendant fields. A possible classification
of these links will also be addressed.
\vskip.5cm
\noindent{\em Acknowledgements}
\vskip.1cm
\noindent The authors thank L.-P. Arguin, P. Jacob,
F. Loranger, P. Mathieu, and Y. Saint-Aubin for discussions, 
and JR thanks M. Bauer, G. Lawler and O. Schramm for helpful correspondence.
The authors are very grateful to S. Chakravarty for pointing out
a mistake in a previous version of this work.

\end{document}